\documentclass[10pt,journal]{IEEEtran}
\usepackage{mathtools}
\usepackage{amssymb,amsmath}
\usepackage[T1]{fontenc}
\usepackage{ifpdf}
\usepackage{url}
\ifpdf
\usepackage[pdftex]{graphicx}
\graphicspath{{figs/}}
\DeclareGraphicsExtensions{.pdf,.png,.jpg}
\else
\usepackage[dvips]{graphicx}
\graphicspath{{eps/}}
\DeclareGraphicsExtensions{.eps}
\fi
\usepackage{subfigure}
\usepackage[export]{adjustbox}
\usepackage{caption}
\usepackage{float}
\usepackage{setspace}
\usepackage{balance}
\begin{document}
\title{Distributed and Managed: \\
Research Challenges and Opportunities of the \\
Next Generation Cyber-Physical Systems}
\author{\IEEEauthorblockN{Gabor Karsai, Daniel Balasubramanian, Abhishek Dubey, and William R. Otte\\}
\IEEEauthorblockA{
  Institute for Software-Integrated Systems, Vanderbilt University,\\
  Nashville, TN 37235, USA \\
  Email:\{gabor,daniel,dabhishe,wotte\}@isis.vanderbilt.edu}
}

\maketitle


\vspace{-0.7in}
\begin{abstract}

 
Cyber-physical systems increasingly rely on distributed computing platforms 
where sensing, computing, actuation, and communication resources 
are shared by a multitude of applications.  
Such `cyber-physical cloud computing platforms' present novel
challenges because the system is built from mobile embedded devices,
is inherently distributed, and typically suffers from highly fluctuating
connectivity among the modules.  Architecting software for these
systems raises many challenges not present in traditional cloud
computing. Effective management of constrained resources and
application isolation without adversely affecting performance are
necessary.  Autonomous fault management and real-time performance
requirements must be met in a verifiable manner.  It is also both
critical and challenging to support multiple end-users whose diverse
software applications have changing demands for computational and
communication resources, while operating on different levels and in
separate domains of security.

The solution presented in this paper is based on a layered
architecture consisting of a novel operating system, a middleware
layer, and component-structured applications. The component model
facilitates the construction of software applications from modular and
reusable components that are deployed in the distributed system and
interact only through well-defined mechanisms. The complexity of
creating applications and performing system integration is mitigated
through the use of a domain-specific model-driven development process
that relies on a domain-specific modeling language and its
accompanying graphical modeling tools, software generators for
synthesizing infrastructure code, and the extensive use of model-based
analysis for verification and validation.

\end{abstract}


\section{Introduction}
\label{sec:intro}

Distributed real-time embedded systems that interact with the physical
world are ubiquitous and pervasive. We are relying on an increasing
number of such systems that provide services to a large number of
users. Fractionated spacecraft (i.e., cluster of satellites)
that performs wide-area sensing of the Earth, swarms of UAVs that
survey storm damage, and the intelligent power devices that are
essential for a `smart' (power) grid are just a few illustrative
examples for this new generation of systems. While distributed and
real-time systems have been built for many decades, there are some
novel properties and requirements for the engineering of such systems
that we need to recognize and address.

First, we have to note that these systems are `cyber-physical', that is,
they interact with the physical world. Hence all software design,
implementation, and verification decisions should be guided by the fact
that physics imposes timing constraints on the computational and
communication activities, and the implementation must obey these
constraints. Furthermore, as the software system may effect changes in
its physical environment these changes must verifiably satisfy safety
requirements for the overall system.

\begin{figure}[ht]
\centering
\includegraphics[width=0.86\columnwidth]{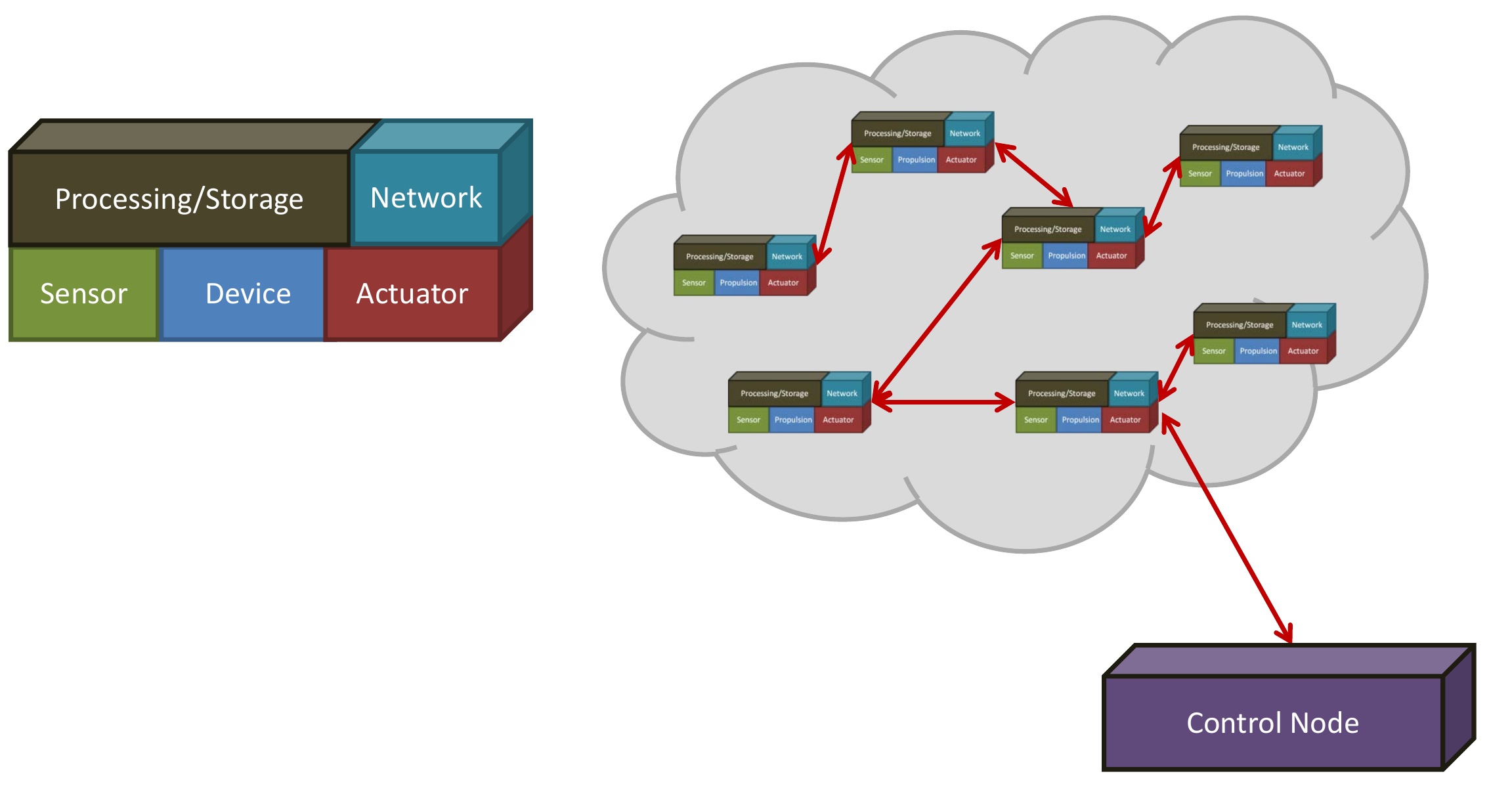}
\caption{Typical node and cloud of nodes.}
\label{fig:cloud_cps}
\end{figure}

Second, we have to understand that these systems are 
\textit{platforms}. That is, they are increasingly built not as a
single use, single function network, but as networked platforms that
can be used by many, possibly concurrent users. The platform is
relatively stable and provide common core services to all applications. 
However,  the applications those run on the platform
change fairly regularly due to software updates or because new applications
have been developed. 
Figure~\ref{fig:cloud_cps} shows a typical node of this
distributed platform on the left, along with a cloud of nodes that are
communicating via a network where at least one of the nodes has a
communication link to a control node. Nodes can join and leave the cloud 
during operation.

Third, these systems are used for \textit{distributed}
applications. Applications typically span multiple nodes, for reasons
related to the availability of resources: some nodes may have sensors,
some may have actuators, some may have the computing or storage
resources, some applications need more than the processing power
available on one node. Therefore, applications that use these resources
have to be architected such that they rely on loosely connected,
interacting components, running on different processors. Applications
can be organically assembled from components that provide specific
services, and components may be used (or re-used) by many active
applications. Obviously, the cluster of computing nodes runs
many applications concurrently.

Fourth, the platform is often a critical resource - possibly a
societal system, whose use must be carefully monitored and controlled
by a responsible owner. Therefore, these systems are \textit{managed} by some authority. Furthermore, as the
platform can be used and shared by many applications, possibly originating from
different organizations, the  platform and thus the system needs to be actively managed 
to avoid `tragedy of the commons'
type failures. Additionally, because of the embedded nature of the system,  deployment and control of applications need to ensure that the 
systems' (often scarce) resources are provisioned. 



Fifth, \textit{security} cannot be an afterthought. Information flows in
general and access to shared resources in particular should be
controlled under some overarching security policy. For instance, high
quality, sensitive customer data (from the electric grid) cannot be
made available to untrusted applications that are supplied by parties
needing access to derived data containing daily averages only -- and
those applications should not have any means to access that high-grade,
sensitive data. Furthermore, applications supplied by users cannot be
trusted, and the platform must protect itself from abuse by such
applications. If multiple applications run on the platform
concurrently, and there is a need for some degree of data sharing
among the applications, the platform must permit that while enforcing
the security policies defined for the system.

Sixth, \textit{resilience} is essential. Anything can go wrong at any
time: faults in the computing and communication hardware, in the
platform, in the application software. Moreover unanticipated changes
in the system (erroneous updates) or in the environment must be
survivable and the system should recover. The system here includes
both the platform, as well as the distributed applications.

Reading the above list one might argue that existing cloud computing
platforms based on virtualization technologies already provide a
solution for all these requirements. However, this is not the case for
the following reasons. Existing cloud computing platforms were not
designed with the requirements of real-time embedded systems, where
operating under resource constraints and timing requirements are
essential. The distributed applications here need to not only scale,
but to also  satisfy timing and security requirements. Interactions
with physical devices (sensor, actuators, special purpose hardware) is
rarely an issue in conventional cloud computing platforms --
everything is virtualized, without consideration for the management of
resources that are part of the system but not the computing
platform. It appears that current cloud computing platforms are not
prepared for mission critical real-time embedded systems, in general.

Arguably, the challenges listed above define a new a category of
systems that is emerging today. In this paper we present some initial
ideas and relevant research questions that will hopefully be addressed
by the research community. The next section discusses the issues of an
overall architecture for such systems. The section following discusses
the needs for a development toolsuite, which is followed by a section
on some initial results. A review of relevant related work is followed
by a summary and conclusions.

\section{Platform architecture}
\label{sec:architecture}
\begin{figure}[t]
\centering \includegraphics[width=0.8\columnwidth]{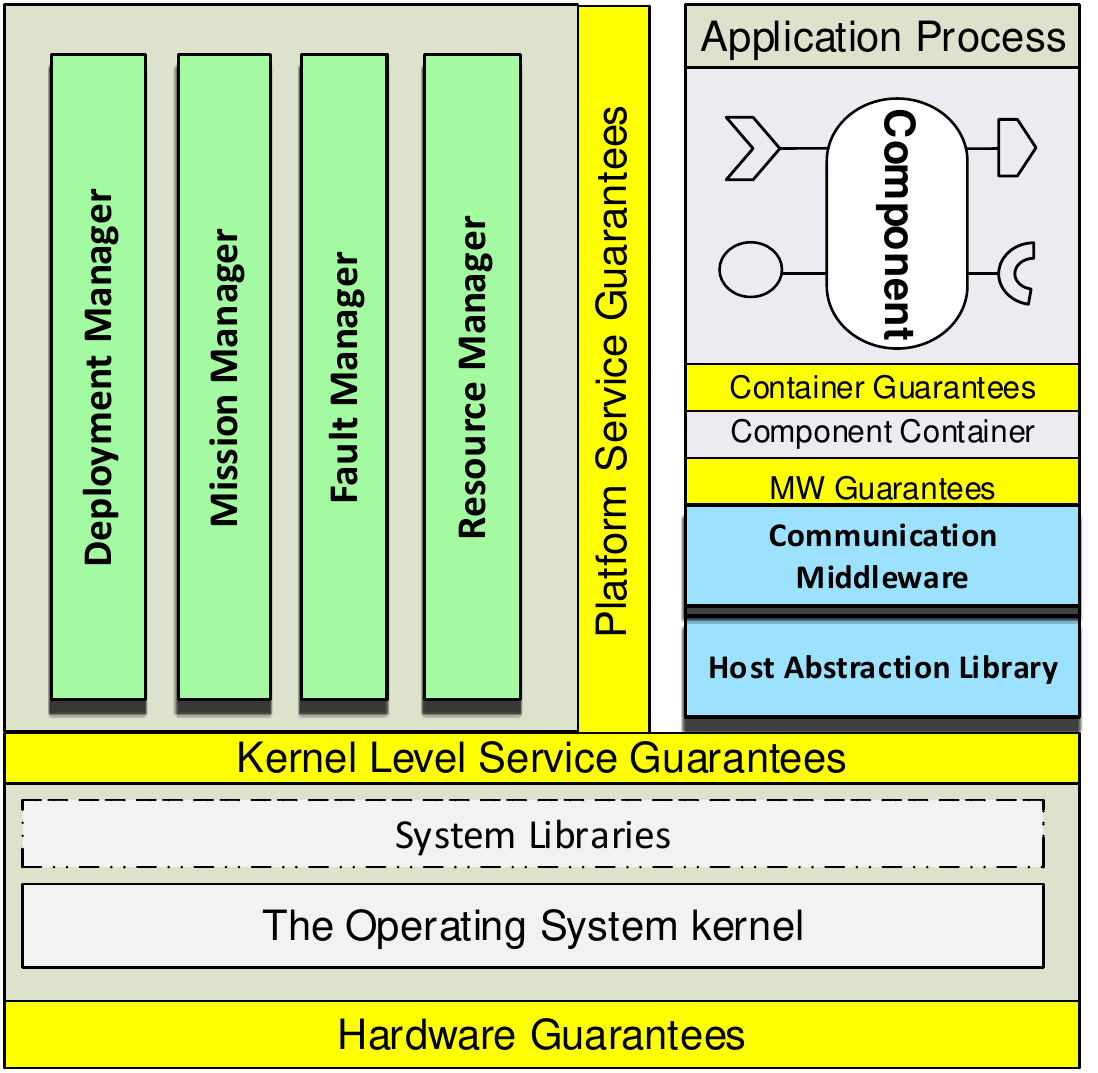}
\caption{Layered architecture of the software platform running in each
  node of the distributed system. The guarantees (assumed and
  provided) are indicated between the layers.}
\label{fig:layered_architecture}
\vspace{-0.2in}
\end{figure}
We aim at building a reusable software platform that can be applied
across many application domains, and many processing and communication
platforms. The software platform should provide solutions to core
resource management problems, support security, and provide services
that are application independent.

This platform  can be built as a multi-layer architecture that 
addresses these issues, as shown on Fig~\ref{fig:layered_architecture}. 
At the lowest level, an operating system kernel provides the core resource sharing and
management functions, as well as the isolation from hardware specific
details. The kernel is typically accessed via `system libraries' that
provide a convenient interface to kernel services. Layered upon this
foundation there is a middleware layer to provide higher-level, reusable
communication (i.e. messaging) and resource management services. The
next layer up provides the component abstractions, in order to support
component-oriented development or distributed applications. The
platform should provide services for application configuration
and lifecycle management (the deployment manager), for application
control (the mission manager), for the handling of faults arising
during operations (the fault manager), and for the management of
resources (the resource manager). Note that by `managers' here we mean
critical, privileged applications that run outside of the core
operating system, and provide complex and long-term management services.

Note that a layered architecture helps with establishing assurances
for the overall systems. At the bottom, the hardware layer provides
guarantees about correct behavior (that was verified by the hardware
vendor). The kernel can assume these and provide its own guarantees to
the higher layers that, in turn, provide their guarantees to the higher 
layers, etc.

\subsection{Platform kernel}

At the lowest layer of the software platform, the kernel  encapsulates device drivers and provides processor scheduling and
networking features, but it also needs to address the real-time,
resilience, and security requirements of the application
domain. Borrowing language from the computer security community, the
kernel has to be part of the Trusted Computing Base (TCB) that
provides guarantees and is built and verified to high-assurance standards
\cite{Rushby84atrusted}.

Real-time requirements can be addressed by a number of
factors. Interrupt latencies (i.e. the worst-case delay elapsed
between the arrival of an interrupt and the release of an activity the
responds to that interrupt) should be bounded and known. System calls
should always be configurably time-bounded, and should return with a 
timeout error in case of unexpected delays to prevent the caller application from 
being unacceptably delayed.  The kernel should support a number of
scheduling policies that provide verifiable guarantees for timeliness
of task execution. Furthermore, it should allow application tasks of
different criticality levels to share the CPU. All tasks of an
application should run at the same criticality level and this should
be reflected in the available scheduling policies.  Because of the
often critical nature of the applications, the scheduling models
provided by the kernel must support timing analysis.

Schedulability analysis is very problematic in the most general case, with
completely unconstrained task behaviors. However, it can become feasible
when restrictions are placed on the behavior of and the interaction
amongst the tasks. As discussed below, an application-level software
\textit{component model} can provide such restrictions, such that the
component-level schedulability becomes manageable. However, the kernel
should be kept simple and provide only core scheduling services that
support potentially many component models. According to experience,
tasks should be able to operate within a shared address space
(i.e. threads) as well as separated address spaces
(i.e. processes). Finally, the kernel scheduler should be able to take
advantage of multi-core architectures and be able to schedule tasks
on different cores, possibly under application control.

Communication links in a distributed system are a critical resource,
especially when they are scarce and highly dynamic, as in mobile
ad-hoc networks. Hence the communication facilities, including the
protocol stack, should be implemented accordingly. As a minimum, the
kernel should support a multitude of transport protocols, preferably
above a common network protocol.

Real-time support must be available for the communications as
well. Datagrams (message blocks) should be time stamped by the kernel
such that message recipients are aware of the message transfer
delays. This necessitates clock synchronization across the nodes;
IEEE or IEEE 802.1AS can serve as the facility to support that. 
Furthermore, the network layer should support time-constrained
real-time communications with guarantees. For some classes of network
traffic existing best-effort approaches (like TCP/IP) are
insufficient, and real-time protocols are needed. The solution here 
necessitates a multitude of network traffic classes, including sporadic but highly critical
traffic, guaranteed-bandwidth time triggered traffic, rate constrained
traffic, and best effort traffic\footnote{\vspace{-0.05in}The standard
SAE AS6802: Time-Triggered Ethernet has similar traffic classes}. The kernel, as the ultimate 
resource manager is to support the sharing of the communication link(s) and 
is to permit applications to  select the traffic classes needed for 
their specific network flows. Furthermore, if the
communication channel is not able to provide the expected performance
anymore the kernel should signal the application so that it can adapt to this change.

Security (i.e. confidentiality, integrity, and authenticity) 
of communications is a critical issue in some of the application domains. 
On the lowest layers, features for secure communication should
be available, possibly supported by the communication hardware
itself and cryptography engines. However, as the
applications running on the platform are not necessarily trusted,
their communication capabilities need to be constrained as
well. Mandatory Access Control (MAC) with Multi Level Security (MLS)
\cite{BellLaPadula} on the network and the messages may be necessary,
in which case the kernel has to provide support for (1) the
trustworthy configuration of network communications and (2) labeled
communications between parties.  The first means that only privileged,
trusted service processes are permitted to configure the network and
the communication flows in the network. We expect that untrusted
processes are not permitted to simply open a communication channel to the network and
talk to any network address -- only trusted service processes can
create the network connections, and once initiated the
communication endpoints are handed over to the untrusted processes for use. The
second means that, following the principles of labeled communications,
each message transmitter and receiver is provided with a label set, by
an external authority. These labels are to be used in each communication
operation by the application, and their correct use is validated and enforced 
by the TCB.  Note that while these technologies have been 
originally developed for government applications, security awareness on a 
shared computing platform necessitates their use.

A communication flow is valid only between parties with labels that
satisfy the rule that information can flow only from lower to
higher or between equal labels (according to the domination relation). 
Assuming an increasing order of sensitivity: \texttt{Confidential} <
\texttt{CompetitionSensitive} < \texttt{ManagementOnly}, 
\emph{e.g.}, a \texttt{CompetitionSensitive} process for mission A can read \texttt{Confidential} or
\texttt{CompetitionSensitive} data for mission A, but not  \texttt{ManagementOnly} data for mission A or
\texttt{CompetitionSensitive} data for mission B.  When the transmitter wishes to transfer a
message, it has to supply a message label that must match with one of
the labels in its own set, and satisfies the MLS rule. The kernel,
which is part of the TCB, performs this check on each message -- both
on the transmitter and the receiver side. This machinery can ensure
that processes always follow the communication constraints defined by
a security policy.

The operating system runs processes; both application and service
processes. To distinguish between these a capability mechanism is
needed that controls what operating system services a process can
use. For example, in order to prevent the unchecked proliferation of
application processes, only privileged processes should be permitted
to create new processes. When a process is created, its parent process 
should specify what capabilities the child has, which can only be a subset of
the capabilities of the parent.

\vspace{-0.05in}
\subsection{Platform services}

As mentioned above, platform services are needed to perform management
functions on the running system that are outside of the scope of
typical applications. Note that platform services perform critical
functions that require privileges, hence the platform services are
part of the TCB. We envision at least four kinds of management
services:

  \textbf{Deployment management:} As stated in the introduction,
  the envisioned systems are managed by some management authority,
  presumably over a network connection. Each node in the system has to
  have a service that can download, install, configure, activate,
  teardown, and remove the distributed applications. This service is
  essentially the top-level configuration manager for the node. Note
  that it itself should be fault-tolerant (i.e. able to manage faults during
  the deployment process), should obey and enforce security policies,
  and should be responsive (per real-time requirements).

 \textbf{Mission management:} Beyond deployment, there is a need
  for a service to manage the execution of applications. One should be
  able to activate and de-activate applications based on triggering
  events or the elapse of time. Triggering events can be generated by
  applications or the services. Mission management should include support
  for system auditing (including logging control) and debugging.

 \textbf{Fault management:} Resilience to faults is a core
  requirement for the system. We envision that the fault management is
  autonomous: the system attempts to restore functionality, if
  possible, without external intervention. Obviously, it may be
  necessary that the system cannot manage a fault on its own, and it
  has to contact its management authority.  While fault management is
  inherently a shared responsibility of all layers (including
  applications), there are some system-level issues that
  can be addressed by a dedicated service. For instance, if an
  application unexpectedly terminates, a fault management reaction
  could be an attempt to restart the application, and if that fails
  then attempt to restart the application on another node, 
  the capability facilitated by a fault management service. Note
  that the software platform is not to define fixed policies for
  fault management (e.g. try restart five times, then re-allocate),
  rather it is to provide mechanisms that allow implementing any such
  policies (e.g. by scripting the behavior of the fault manager
  service).

 \textbf{Resource management:} Embedded systems are typically
  resource constrained, hence unbounded resource usage cannot be
  permitted. This can be strictly managed by a static quota system,
  where developers declare the resource needs of their applications,
  then a system integrator verifies that such resource needs are
  acceptable (i.e. the application is `admissible'), and then the
  software platform enforces these quotas. If the application attempts
  to obtain more resources than it was declared, the request will fail
  (and the application has to handle this failure). This method is too
  strict, however, and may use resources very inefficiently. A
  resource manager service can implement a more complex, dynamic
  resource allocation policy, where applications can dynamically
  request and release resources, and the service honors or rejects
  these requests while maximizing system utility. Note that a critical
  resource is network bandwidth (if it is limited and/or fluctuating),
  and the dynamic management of communication bandwidth that maximizes
  system utility is a challenge.

\subsection{Middleware}
All modern distributed software systems are built using middleware
libraries that provide core communication abstractions for
object-based systems. These abstractions are to facilitate
prototypical component interactions. Industry standards and
pragmatic experience shows that a well-defined, small set of 
interaction patterns can provide a solid foundation for building
applications. The set includes: (1) Point-to-point interactions when
when an object wants to invoke specific services of another object. 
The interaction can be synchronous (call-return) or  asynchronous (call-callback). 
Note that the client and server are coupled and are involved in bi-directional messaging.
(2) Publish-subscribe interactions when a publisher generates data samples, 
which are then asynchronously consumed by interested subscribers. Note that the
publishers are loosely coupled, and not directly known to each other. 
While additional, more complex interactions may also be needed, the interactions
should be facilitated in conjunction with overall system requirements.
For instance, the interactions can be subject to timing constraints, and the scheduling
of the message exchanges should be done accordingly. The interactions
are also subject to the security policies - only permitted information
flows can be utilized to facilitate an interaction. The interactions
have to be implemented in conjunction with the fault management
architecture: objects participating in the interactions must become
aware of faults (originating from the network, for instance), and
should be able to rely on fault tolerant services, if available.

\vspace{-0.05in}
\subsection{Component model}

We envision a component-oriented software development for the
platform. Obviously, this necessitates a precisely defined abstract
component model that helps developers to build robust systems from
reusable components. The implementation of the component model 
must rely on a robust component framework that
facilitates and mediates all interactions among the components. The
component model should clearly define how component activities are
scheduled, based on events or the elapse of time, and what the
component lifecycle is.

The component model is subject to all requirements mentioned above. It
has to support real-time requirements: we want to be able to predict
the timing properties of the system based on the timing properties of
the components and their specific interactions. The component model
should support security policies, and should provide for
fault management, including anomaly detection, diagnosis, and fault
mitigation.

\section{Development tools}
\label{sec:development}

Developing code for modern software platforms (e.g. Android or iOS)
cannot be done without sophisticated tool support. This issue is
compounded by the complexity of distributed systems, where not only
code development has to be done, but also complex configuration and
allocation decisions have to be made and implemented.

As discussed above, the platform supports a component model with
complex interaction semantics. Coding for such a component model by
hand is quite difficult and error prone, hence higher-level
abstractions, such as models, are needed. There is a need for a
modeling language specific for the component model that is easy to use
and mitigates accidental complexity. Furthermore, the modeling
language should facilitate the composition of applications from
components. As we are building a distributed system, the models
should also encompass the (static or dynamic) configuration of the
network with computing nodes and communication links. Many
cross-cutting aspects, like resource quotas, fault management
strategies, security labels for secure communications, etc. should
also be represented. Finally, the allocation of applications to computing
nodes and information flows to network links should also be modeled,
either explicitly (to support static allocation) or implicitly (to
support dynamic allocation).

In summary, we envision a wide-spectrum domain-specific modeling
language that covers all of the above areas. General purpose modeling
languages (e.g. UML) or their specializations (e.g. MARTE) solve only
part of the problem, and often in a somewhat cumbersome way,
e.g. using stereotypes. Arguably, a dedicated, platform-specific modeling
language is a better approach.

The development toolchain should be able to support both conventional
(code-oriented) and model-based development of software
components. The first one is needed for general purpose components,
while the second one opens up the opportunity to use the results of
model-based development tools (like Simulink/Stateflow). Tool
integration to ensure semantic interoperability across development
tools is essential.

Finally, the development tools should include tools for checking the
correctness of the modeled applications and analyzing system
properties including schedulability and the ability to compare
alternative deployment strategies.

\section{Preliminary results}
\label{sec:results}

DREMS\footnote{\vspace{-0.1in}http://www.isis.vanderbilt.edu/drems} is a software
infrastructure for designing, implementing, configuring, deploying and
managing distributed real-time embedded systems that consists of two
major subsystems: (1) a design-time toolsuite for modeling, analysis,
synthesis, implementation, debugging, testing, and maintenance of
application software built from reusable components, and (2) a
run-time software platform for deploying, managing, and operating
application software on a network of computing nodes. The platform is
tailored towards a managed network of computers and distributed
software applications running on that network of nodes, i.e. a cluster.

The toolsuite supports a model-based paradigm of software development
for distributed, real-time, embedded systems where modeling tools and
generators automate the tedious parts of software development and also
provide a design-time framework for the analysis of software
systems. The run-time software platform reduces the complexity and
increases the reliability of software applications by providing
reusable technological building blocks in the form of an operating
system, middleware, and application management services.

\subsection{DREMS Architecture}

DREMS is a complete, end-to-end solution for software development:
from modeling tools to code to deployed applications. It is open and
extensible, and relies on open industry standards, well-tested
functionality and high-performance tools. It focuses on the
architectural issues of the software, and promotes the modeling of
application software, where the models are directly used in the
construction of the software.

Software applications running on the DREMS platform are distributed:
an application consists of one or more \textit{actors} that run in parallel,
typically on different nodes of a network. Actors specialize the
concept of processes: they have identity with state, can be
migrated from node to node.  Actors are created,
deployed, configured, and managed by a special service of the run-time
platform: the deployment manager - a privileged, distributed, and
fault tolerant actor, present on each node of the system, that
performs all management functions for application actors. An actor can
also be assigned a set of limited resources of the node on which it runs:
memory and file space, a share of CPU time, and a share of the network
bandwidth.

Applications are built from software components - hosted by actors -
that only  interact via  well-defined interaction patterns using
security-labeled messages, and are allowed to use specific sets of
services provided by the operating system, including messaging and 
thread synchronization operations. Note that components use these 
indirectly: via the middleware.

The middleware libraries implement the high-level communication
abstractions: synchronous and asynchronous interactions, on top of the
low-level services provided by the underlying distributed hardware
platform. Interaction patterns include (1) point-to-point interactions
(in the form of synchronous and asynchronous remote method
invocations), and (2) group communications (in the form of
anonymous publish-subscribe interactions). Component operations can
be event-driven or time-triggered, enabling time-driven
applications. Message exchanges via the low-level messaging services
are time-stamped, thus message receivers are aware of when the message
was sent. Hence temporal ordering of events can be established
(assuming the clocks of the computing nodes are synchronized).

Specialized, verified platform actors provide system-wide high-level
services: application deployment, fault management, controlled access
to I/O devices, etc. Each application actor exposes the interface(s)
of one or more of its components that the components of applications
can interact with using the same interaction patterns. Applications
can also interact with each other the same way: exposed interfaces and
precisely defined interaction patterns.

The DREMS Operating System - a set of extensions to the Linux kernel -
implements all the critical low-level services to support resource
sharing (including spatial and temporal partitioning), actor management,
secure (labeled and managed) information flows, and fault tolerance. A
key feature of the OS layer is support for temporal partitions
(similar to the ARINC-653 standard): actors can be assigned to a
fixed duration, periodically repeating interval of the CPU's time so
that they have a guaranteed access to the processor in that
interval. In other words, the actors can have an assured bandwidth to
utilize the CPU and actors in separate temporal partitions cannot
inadvertently interfere with each other via the CPU.

\vspace{-0.04in}
\subsection{Run-time Software Platform}

The implementation of the run-time software platform has several layers. 
Practically all layers are based on existing and proven open-source
technology. Starting from the bottom, the operating system layer
extends the Linux kernel with a number of specific services, but it
strongly relies on the code available in the Linux kernel (currently:
version 3.2.17). This permits the use of DREMS services for the actors,
but also keeps the Linux system calls for debugging and monitoring 
purposes. These extensions are in the form of 120+ new system
calls.

The C and C++ run-time support libraries (based on uClibc\footnote{\url{www.uclibc.org}}
and libstdcpp\footnote{\url{http://gcc.gnu.org/libstdc++/}} 
implement the conventional
support services needed by the typical C and C++ programs. The C
run-time library has entry points to access the DREMS OS system
calls. These calls utilize data structures that have been defined
using the standard OMG Interface Definition Language (IDL), and can be
created and manipulated using generated constructor and manipulation
operators. The implementation of the DREMS operating system calls
checks the integrity of all data structures passed on the
interface. This enables validation of the data structures on the
interface, preventing potential abuse of the system calls.

Layered on the C and C++ run-time libraries the Adaptive Communication
Environment (ACE) libraries provide a low-overhead isolation layer for
the higher level middleware elements that support CORBA and DDS. The
CORBA implementation is based on The ACE ORB (TAO, currently: version
6.1.4) that implements a subset of the CORBA standard for facilitating
point-to-point interactions between distributed objects. Such
interactions are in the form of Remote Method Invocations (RMIs) or
Asynchronous Method Invocations (AMIs). RMIs follow the call-return
semantics, where the caller waits until the server responds, while the
AMIs follow the call-return-callback semantics, where the caller
continues immediately and the response from the server is handled by a
registered callback operation of the client. The CORBA subset
implemented by the middleware has been selected to support a minimal
set of core functions that are suitable for resource-constrained
embedded systems.  The DDS implementation is based on the OpenDDS
(currently: version 3.4) that implements a subset of the DDS standard
for facilitating anonymous publish/subscribe interactions among
distributed objects. 
There are several quality-of-service attributes
associated with publishers and subscribers that control features like
buffering, reliability, delivery rate, etc. DDS is designed to be
highly scalable, and its implementations meet the requirements of
mission-critical applications.

CORBA and DDS provide for data exchange and basic interactions between
distributed objects, but in DREMS objects are packaged into
higher-level units called components. A component
\cite{ISIS_F6_ISORC:13} publishes and subscribes to various topics
(possibly many), implements (provides) interface(s), and expects
(requires) implementations of other interfaces. Note that a component
may contain several, tightly coupled objects.  Components may expose
parts of their observable state via read-only state variables,
accessible through specific methods. Components are configured via
configurable parameters. Their operations are
scheduled based on events or elapse of time. An event can be the
arrival of a message the component has subscribed to or an incoming
request on a provided interface. Time triggering is done by
associating a timer with the component that invokes a selected
operation on the component when a set amount of time elapses,
possibly periodically repeating the operation. Component operations
can perform computations, publish messages, and call out to other
components via the required interfaces. To avoid having to write
complex locking and synchronization logic for components, component
operations are always single threaded: inside of one component at most
one thread can be active at any time.  Actors are formed from
interacting components, and applications are formed from actors that
interact with each other via their interacting components. Actors
(together with their components) can be deployed on different nodes of
a network, but their composition and interactions are always clearly
defined: they must happen either via remote method invocations or via
publish/subscribe interactions.

\begin{figure}[t]
\centering \includegraphics[width=0.9\columnwidth]{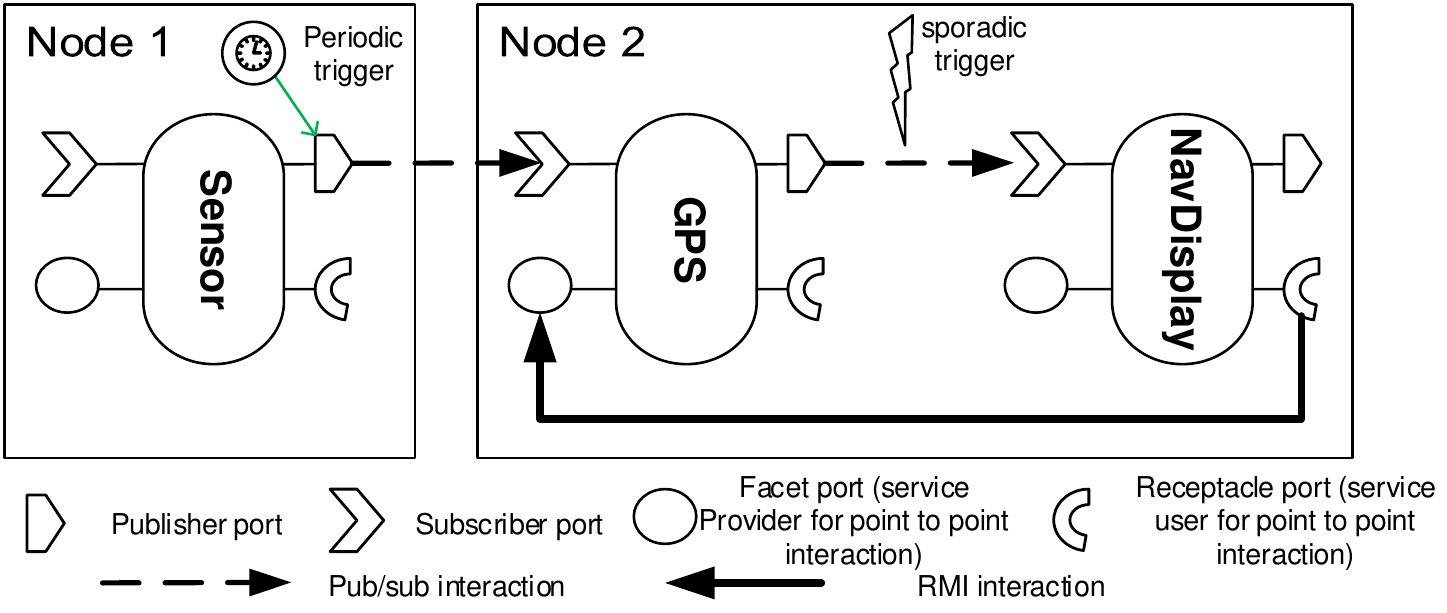}
\caption{Component-based distributed system example}
\label{fig:component_architecture}
\vspace{-0.1in}
\end{figure}

Figure~\ref{fig:component_architecture} shows an application where a
Sensor component periodically (P) publishes a message that a GPS
component subscribes to, and which, in turn, sporadically (S) publishes
another message that a NAVDisplay component consumes. This last
component invokes the GPS component via a provided interface, when it
needs to refresh its own state. The messages published can be quite
small, while the method invocation (that happens less frequently, and
on demand) may transfer larger amounts of data. 


The run-time software platform includes a key platform actor: the
Deployment Manager (DM) that instantiates, configures, activates, deactivates,
and dismantles applications. Every node on a network has a copy of the DM that acts
as a controller for all applications on that node. The DMs communicate
with each other, with one being the lead  `cluster' DM. This, cluster
leader orchestrates the deployment of applications across cluster
with the help of the node DMs. For deployment, the binaries of
application components should be installed on each node, then the
cluster lead DM is provided a deployment plan that is generated from
application models and executes the plan, coordinating the activities
of node level DMs which start the actors, installs components,
configures the network connections among the components, etc., and
finally activates the components. This last step releases the
execution threads of the components. When the applications need to be
removed, the DM stops the components, withdraws the network
configuration, and stops the actors. A key feature of the deployment
process is that the network connections among the parts: i.e. actors
and components of the distributed application are managed: the
application business logic does not have to deal with this problem;
everything is set up based on the deployment plan.

\subsection{Design-time Development Platform}
Configuring the middleware and writing code that takes advantage of
the component framework is a highly non-trivial and tedious task. To
mitigate this problem and to enable programmer productivity, a
model-driven development environment is available that simplifies the
tasks of the application developers and system integrators.

In this environment, developers define via graphical and textual
models various properties of the application, including: interface and
message types, components types (in terms interfaces and
publish/subscribe message types), component implementations, component
assemblies, and applications (in terms interacting components and
actors containing them). Additionally, the hardware platform for the
cluster are modeled: processors, network and device interfaces,
network addresses, etc. Finally, the deployment of the application(s)
on the hardware platform are also modeled, as the mapping of actors onto
hardware nodes, and information flows onto network links. 
Models are processed by code generators
that produce several artifacts from them: source code, configuration
files, build system artifacts that facilitate the automated
compilation and linking of the components, and other documents. The
application developer is expected to provide the component
implementation code in the form of C++ code (currently, in the future:
any other, supported executable language) and add it to the generated
code. The compilation and debugging of the applications can happen
with the help of a conventional development environment
(currently: Eclipse) that supports editing, compiling, and debugging
the code.  The result of this process is a set of component
executables and the deployment plan - ready to be deployed on a
cluster of nodes.

The model-driven approach has several benefits. (1) The model serves
as the single source of all structural and configuration information
for the system. (2) The tedious work of crafting middleware `glue'
code and configuration files for deployment is automated: everything
is derived programmatically from the models. (3) The models provide an
explicit representation of the architecture of all the applications
running on the system - this enables architectural and performance
analysis on the system before it is executed. (4) Models can also be
used for rapidly creating `mockup' components and applications for
rapid prototyping and evaluation.

\subsection{Example: Cluster flight control and sensor processing}

We have evaluated the DREMS prototype on several examples. The graphical modeling
tool runs on Windows, the code development and cross-compilation tools on a Linux platform, 
while DREMS is on a set of networked embedded x86-based devices (3 iBX-530 industrial computers). 
Deployment and configuration is done from the Linux machine, via the network. 
Several small scale tests were used to validate that the platform is functional. A more
realistic application involved a distributed flight control software applications (2 actors on each node,
with 3-4 components each), and a sensor processing application (dissimilar actors on each node). The flight control
actors share critical, but low-bandwidth data, the sensor application shares high bandwidth, but low criticality data. The two sets of applications run in different security domains, and in different temporal partitions. We were scheduling the applications in partitions of 100 msec duration, and were experimenting with variable bandwidth between the nodes. All designed and implemented features were functional, including component interactions, partition scheduling, security labeling and information flow separation, application deployment and control. The applications have been constructed using the model-driven development toolchain; the model had about 100 distinct elements. Component code was hand inserted into the skeleton code generated by the software generators, followed by compilation using Eclipse with a cross-compiler for the platform. The model-driven generation produced all the infrastructure code, simplifying the task of the developer. A detailed report of this experiments can be found in \cite{drems-rtss-2013}.

\section{Related work}
\label{sec:rel_work}


There are several architecture description languages for embedded systems, including the Architecture Analysis and Design Language (AADL) \cite{AADL_Intro:06}, SysML ~\cite{SysML_v1.3:12} and the Modeling and Analysis of Realtime and Embedded (MARTE) ~\cite{MARTE_v1.1:11} systems profile for UML. These are general purpose modeling languages that can be used across a wide variety of systems. Because of specific features that are tightly integrated into our system, such as security labels and partition scheduling, we designed a dedicated domain-specific modeling language to describe DREMS systems and applications. However, automated transformation from our modeling language to these general purpose languages is possible and may be used to leverage some of their analysis capabilities.

A similar toolsuite that also uses a domain-specific approach for component-based systems is described in ~\cite{Metamodel_ECFMA:10}. That work focuses primarily on support for highly dynamic environments that require adaptation, and hence their environment supports dynamic updating and reconfiguration of models based on feedback from the running system. The biggest differences between that work and DREMS are that DREMS supports multiple messaging semantics and has built-in support for security in both the kernel and middleware layers.

Our previous work in modeling component-based systems includes the CoSMIC ~\cite{gokhale2008model, Schmidt:05e} tool suite, which assists with the model-based development, configuration and deployment of CORBA Component Model-based applications. While DREMS is more extensive than CoSMIC and provides the ability to model elements like hardware and task schedules, experience from the CoSMIC project helped guide certain design aspects of component modeling inside DREMS.

The ARINC-653 Component Model (ACM)~\cite{ACM_SPE:10}, which implements a component model for the ARINC-653 standard~\cite{ARINC-653} for avionics computing, forms the basis for the DREMS component model. DREMS extends the temporal partitioning scheduling method used by ACM by allowing multiple actors (processes) per temporal partition, a valuable feature for components that interact through synchronous messages. Further, the DREMS component model is designed to promote deadlock/race condition-free behavior in components.

The secure transport feature of DREMS is based on multi-level security (MLS) \cite{BellLaPadula}. All messages have a security label and must obey a set of mandatory access control (MAC) policies. The main novelty in DREMS with respect to MLS is the concept of multi-domain labels \cite{OlinSibertLabels} to support secure communication among actors from different organizations.

A more detailed description of the system requirements for DREMS and design principles used to meet those requirements is available in \cite{DREMS13Software}.
\section{Conclusions}
\label{sec:conclusions}

DREMS is a prototype, end-to-end solution for building and running
distributed real-time embedded applications. It contains not only a
run-time framework with a state-of-the-art operating system extended
with special features for resource, application, and network
management together with a component framework with a precisely defined
model of computation, but also a model-driven development toolchain
that assists developers and integrators in managing the development
process.

But DREMS is only a \textit{partial} prototype for a class of software
platforms outlined in the first two sections. We believe that such
software platforms are essential for implementing the next generation
of distributed real-time embedded systems. Embedded systems are not
black boxes anymore, but rather platforms with an evolving and
dynamically changing software ecosystem.

%

\section*{Acknowledgments} 
This work was supported in part by the DARPA System F6 Program
under contract NNA11AC08C. Any opinions, findings, and
conclusions or recommendations expressed in this material
are those of the author(s) and do not  reflect
the views of DARPA. This work was also supported in part by FORCES 
(Foundations Of Resilient CybEr-Physical Systems), 
which receives support from the National Science Foundation.

\balance
\begin{spacing}{0.88}
\bibliographystyle{IEEEtran}
\bibliography{f6}

\begin{thebibliography}{10}
\providecommand{\url}[1]{#1}
\csname url@samestyle\endcsname
\providecommand{\newblock}{\relax}
\providecommand{\bibinfo}[2]{#2}
\providecommand{\BIBentrySTDinterwordspacing}{\spaceskip=0pt\relax}
\providecommand{\BIBentryALTinterwordstretchfactor}{4}
\providecommand{\BIBentryALTinterwordspacing}{\spaceskip=\fontdimen2\font plus
\BIBentryALTinterwordstretchfactor\fontdimen3\font minus
  \fontdimen4\font\relax}
\providecommand{\BIBforeignlanguage}[2]{{%
\expandafter\ifx\csname l@#1\endcsname\relax
\typeout{** WARNING: IEEEtran.bst: No hyphenation pattern has been}%
\typeout{** loaded for the language `#1'. Using the pattern for}%
\typeout{** the default language instead.}%
\else
\language=\csname l@#1\endcsname
\fi
#2}}
\providecommand{\BIBdecl}{\relax}
\BIBdecl

\bibitem{Rushby84atrusted}
J.~Rushby, ``A trusted computing base for embedded systems,'' in
  \emph{Proceedings of the 7th Department of Defense/NBS Computer Security
  Conference}, 1984, pp. 294--311.

\bibitem{BellLaPadula}
D.~E. Bell and L.~J. LaPadula, ``Secure computer systems: Mathematical
  foundations,'' MITRE, Technical Report 2547, Volume I, 1973.

\bibitem{ISIS_F6_ISORC:13}
W.~R. Otte, A.~Dubey, S.~Pradhan, P.~Patil, A.~Gokhale, G.~Karsai, and
  J.~Willemsen, ``{F6COM: A Component Model for Resource-Constrained and
  Dynamic Space-Based Computing Environment},'' in \emph{Proceedings of the
  16th IEEE International Symposium on Object-oriented Real-time Distributed
  Computing (ISORC '13)}, Paderborn, Germany, Jun. 2013.

\bibitem{drems-rtss-2013}
W.~Emfinger, P.~Kumar, A.~Dubey, W.~Otte, A.~Gokhale, and G.~Karsai, ``Drems: A
  toolchain and platform for the rapid application development, integration,
  and deployment of managed distributed real-time embedded systems,'' in
  \emph{IEEE Real-time Systems Symposium,}, 2013, {RTSS\@Work}.

\bibitem{AADL_Intro:06}
P.~H. Feiler, D.~P. Gluch, and J.~J. Hudak, ``{The Architecture Analysis \&
  Design Language (AADL): An Introduction},'' DTIC Document, Tech. Rep.
  ADA455842, 2006.

\bibitem{SysML_v1.3:12}
{Object Management Group}, \emph{{Systems Modeling Language (SysML), Version
  1.3}}, {OMG Document formal/2012-06-01}~ed., {Object Management Group}, Jun.
  2012.

\bibitem{MARTE_v1.1:11}
------, \emph{{UML Profile for MARTE: Modeling And Analysis of Real-Time
  Embedded Systems, Version 1.1}}, {OMG Document formal/2011-06-02}~ed.,
  {Object Management Group}, Jun. 2011.

\bibitem{Metamodel_ECFMA:10}
G.~Batori, Z.~Theisz, and D.~Asztalos, ``{Metamodel-based Methodology for
  Dynamic Component Systems},'' \emph{Modelling Foundations and Applications},
  vol. 7349, pp. 275--286, 2012.

\bibitem{gokhale2008model}
A.~Gokhale, K.~Balasubramanian, A.~Krishna, J.~Balasubramanian, G.~Edwards,
  G.~Deng, E.~Turkay, J.~Parsons, and D.~Schmidt, ``{Model driven middleware: A
  new paradigm for developing distributed real-time and embedded systems},''
  \emph{Science of Computer programming}, vol.~73, no.~1, pp. 39--58, 2008.

\bibitem{Schmidt:05e}
K.~Balasubramanian, A.~S. Krishna, E.~Turkay, J.~Balasubramanian, J.~Parsons,
  A.~Gokhale, and D.~C. Schmidt, ``{Applying Model-Driven Development to
  Distributed Real-time and Embedded Avionics Systems},'' \emph{International
  Journal of Embedded Systems: Special Issue on the Design and Verification of
  Real-Time Embedded Software}, vol.~2, pp. 142--155, 2006.

\bibitem{ACM_SPE:10}
A.~Dubey, G.~Karsai, and N.~Mahadevan, ``{A Component Model for Hard Real-time
  Systems: \textsc{CCM} with \textsc{ARINC-653}},'' \emph{Software: Practice
  and Experience}, vol.~41, no.~12, pp. 1517--1550, 2011.

\bibitem{ARINC-653}
\emph{{Document No. 653: Avionics Application Software Standard Inteface (Draft
  15)}}, {ARINC Incorporated}, Annapolis, Maryland, USA, Jan. 1997.

\bibitem{OlinSibertLabels}
O.~Sibert, ``Multiple-domain labels,'' 2011, presented at the F6 Security
  Kickoff.

\bibitem{DREMS13Software}
T.~Levendovszky, A.~Dubey, W.~Otte, D.~Balasubramanian, A.~Coglio, S.~Nyako,
  W.~Emfinger, P.~Kumar, A.~Gokhale, and G.~Karsai, ``Drems: A model-driven
  distributed secure information architecture platform for managed embedded
  systems,'' \emph{IEEE Software}, vol.~99, no. PrePrints, p.~1, 2013.

\end{thebibliography}
\end{spacing}

\end{document}